\documentstyle[prl,aps,twocolumn]{revtex}

\newcommand{\beq}{\begin{eqnarray}}
\newcommand{\eeq}{\end{eqnarray}}

\newcommand{\eps}{\epsilon}

\newcommand{\btem}{\bibitem}
\begin{document}
\preprint{RYUTHP-97/4, Oct. 1997}
\draft

\title{Renormalization-group Resummation of Divergent Series of 
the Perturbative Wave Function of Quantum Anharmonic Oscillator}

\author{Teiji Kunihiro}	

\address{Faculty of Science and Technology, Ryukoku University,
 Seta, Ohtsu, 520-21, Japan}

\date{\today}
 
\maketitle

\begin{abstract}
The renormalization group method  
 is applied  
for  obtaining the  asymptotic form of the wave function of 
the quantum anharmonic oscillator by resumming the 
  perturbation series.
It is shown  that the resumed series is the {\em cumulant} 
 of the naive perturbation series.
Working out up to the sixth order and performing a further resummation
 proposed by Bender and Bettencourt,
  we  find that the agreement with the WKB result becomes worse  
 in the higher orders than the fourth at which the agreement is the 
best.
\end{abstract}

\pacs{PACS numbers: 03.65.Ge,02.30.Mv,11.15.Bt,11.15.Tk}
\narrowtext
It is well known that naive perturbation series 
 are divergent or at best asymptotic series\cite{bo}.
 One needs to resum the divergent series to obtain a
 sensible  result from the perturbation theory.
 Indeed, various resummation techniques  have been devised\cite{bo}.
Recently, 
a unified and mechanical method for global and asymptotic analysis
 has been  proposed by Goldenfeld et al\cite{cgo}.  This is called the
 renormalization group (RG) method. 
 The unique feature  of their method is to start with the naive 
 perturbation theory and allow  secular terms to appear in contrast 
with all previous methods\cite{bo}; adding unperturbed solutions to the 
 perturbed solutions so that the secular terms
 vanish at a "renormalization point" $t=t_0$ and then  applying the
 RG equation, one obtains a resummed perturbation series.
 
Subsequently, the present author formulated the RG method geometrically 
on the  basis of the classical theory of envelopes\cite{kuni};
it was  indicated  that 
the RG equation  $\grave{a}$ la Gell-Mann-Low in the field theory 
can be identified  as  the envelope equation.

 The purpose of the present work is to apply the RG method  as 
 formulated in \cite{kuni}
 to Schr\"{o}dinger equation of the quantum anharmonic oscillator (AHO)
 and obtain the 
 asymptotic form of the wave function.
 
AHO  is a theoretical laboratory for
 examining the validity of various approximation 
techniques\cite{rev1,rev2}.
Recently, Bender and Bettencourt \cite{bb} have shown that  multiple-scale
 perturbation theory (MSPT) can be successfully 
 applied to the quantum anharmonic oscillator; MSPT or the reductive
 perturbation theory is known to be one
 of the  most general method in applied mathematics \cite{bo}
 apart from the RG method for improving perturbative 
expansions\cite{frasca1}. 
They examined the Heisenberg operator equation and the 
 (time-independent) Schr\"{o}dinger equation: The exact closed-form
 solution was found for the Heisenberg equation, and the asymptotic
 behavior of the wave function $\psi (x)$ for large $x$ was
 constructed which agrees with  the WKB result. 
One should remark  here that  a further resummation had to be  adopted 
 for the latter case, which is not intrinsic in 
MSPT and a similar method had been proposed by Ginsburg and Montroll
\cite{gm}. 

Actually the RG method a l\`{a} Goldenfeld et al has been already applied
 to quantum mechanics by others\cite{egus,frasca2}: Egusquiza and
 Valle Basagoiti \cite{egus} applied it to solve the time-dependent
 Heisenberg equation  considered in \cite{bb}, 
while Frasca used it  to solve
 the  time-dependent Schr\"{o}dinger equation  of a two-level
 system. There has been, however, no attempt to apply the RG method to
 obtain asymptotic forms of wave functions. The reason may be
 that it is not trivial to identify the secular terms for the
 wave functions which can be made to vanish at a ``renormalization 
 point'' $x=x_0$. In the present work,  such secular terms
 are successfully identified for the ground and the first excited 
states. 

 We  shall show that the resummation of the
 perturbation series of the  wave functions is performed in 
 the RG method more
 mechanically and explicitly than in MSPT:
The polynomials $f_n(x)$ in the resumed series are explicitly 
given in terms of the
 polynomials $P_n(x)$ obtained in the naive 
perturbation theory\cite{bb}.
Furthermore, it will be found that 
 $\sum_{n=0}^{\infty}\eps^nf_n(x)$ is
 the {\em cumulant} series\cite{cumulant} 
 of the naive perturbation series $\sum_{n=0}^{\infty}\eps^nP_n(x)$.
Since our method is mechanical and easy to perform,
we shall work out up to the sixth order in the perturbative
 expansion 
 and examine how the results of Bender and Bettencourt persist
 or are modified in the higher orders.

Our Hamiltonian for the anharmonic oscillator is given by\cite{notation}
\beq
H=p^2 + \frac{1}{4}x^2 + \frac {1}{4}\eps x^4,
\eeq
and we consider the following Schr\"{o}dinger equation
\beq
(H- E)\psi(x)=0,
\eeq
with the boundary condition
 $\psi(\pm \infty)=0$.
We shall confine ourselves to the ground state for the moment. 
WKB analysis shows that for large $x$, 
\beq
\psi(x)\sim {\rm exp}\{-\sqrt{\eps}\vert x\vert ^3/6\}.
\eeq
We shall examine how the perturbation theory can reproduce the
 WKB behavior or not, as was done in \cite{bb}.
  
\newcommand{\ee}{e^{-x^2/4}}
As preliminaries,
 we first  apply the Bender-Wu method \cite{bw} for 
 performing Rayleigh-Schr\"{o}dinger(RS) perturbative expansion.
The wave function and the eigenvalue are both expanded as power
 series of $\eps$;
\beq
\psi(x) \sim \sum_{n=0}^{\infty}\eps^n y_n(x)\ \ \ {\rm and}
\ \ \ E(\eps) \sim \sum_{n=0}^{\infty}\eps^n E_n.
\eeq
We take the boundary values hence the normalization as;
\beq
y_0(0)=1\ \ \ {\rm and}\ \ \ y_{n\geq 1}(0)=0.
\eeq
The lowest order solution reads
\beq
y_0(x)=e^{-x^2/4}\ \ \ {\rm and} \ \ \ E_0=1/2.
\eeq
The higher order terms with $n\geq 1$ are written as
\beq
 y_n(x)=\ee P_n(x),
\eeq
where $P_n(x)$ is a polynomial. It is readily shown that the polynomials 
satisfy the recursion relation;
\beq
P_n''(x) -x P_n'(x)=\frac{x^4}{4}P_{n-1}(x) - \sum _{j=0}^{n-1}P_j(x)E_{n-j}.
\eeq
This equation determines the polynomials and the eigenvalues $E_n$
 successively. Here it should be remarked that
 $y_n(x)$ in Eq.(7) may be identified as a secular term because
 it is a product of the unperturbed solution and a function that
  increases as $x$ goes large.

The eigenvalues $E_n$ are given by the 
condition (solvability condition)
$\int _{-\infty}^{\infty}dx\ y_0(x)\hat{h}F(x)=0,$ 
where $\hat{h}=(d/dx)^2 -xd/dx$ and $F(x)$ is an arbitrary function
 with which the integral converges. Thus one finds
\beq
E_n=\frac{1}{2\sqrt{\pi}}\int_{-\infty}^{\infty}dx \ y_0(x)
\{
\frac{x^4}{4}P_{n-1}(x) - \sum _{j=1}^{n-1}P_j(x)E_{n-j}
\}.
\eeq
Note that $E_n$ is determined in terms  of  only the polynomials
 $P_j(x)$ with $j\leq n-1$. 
  
With these eigenvalues $E_j$ \ ($j\leq n$), 
the polynomial $P_n(x)$  is determined by Eq.(8).
The general form of the polynomial is expressed as\cite{bb}
\beq
P_n(x)=\sum_{k=1}^{2n}C_{n,k}(-\frac{x^2}{2})^k,
\eeq
where the coefficient $C_{n,k}$ satisfies a recursion relation
\beq
2k C_{n,k}+ C_{n-1, k-2}& =& -(k+1)(2k+1)C_{n, k+1}\nonumber \\ 
 \ \ \ & & + \sum_{j=1}^{n-1}C_{j,k}C_{n-j, 1},
\eeq
 with $C_{n,1}=E_n$. The recursion relation is solved for low $k$ with
 given $n$.
 The polynomials $P_n(x)$ up to the six order are  
 presented in \cite{bb}, which we refer to.


Now we apply the renormalization group method to resum the 
 perturbation series obtained above.
 We shall present the method so that it becomes clear 
 that the notion of envelopes is intrinsically related to the 
method\cite{kuni}:
We shall also make it clear that the RG method concerns
 with the boundary conditions in conformity with the general
 property of the RG methods as emphasized by Shirkov\cite{shirkov}.

First we try to  obtain the wave function $\psi(x;x_0)$
around an initial point $x=x_0$ in a perturbative way:
\beq
\psi(x;x_0)\sim 
\sum_{n=0}^{\infty}\eps^n z_n(x; x_0)\ \ \ {\rm and}
\ \ \ E(\eps) \sim \sum_{n=0}^{\infty}\eps^n E_n,
\eeq
with  the initial or boundary condition (BC) at $x=x_0$;
\beq
\psi(x_0; x_0)= W(x_0).
\eeq
We suppose that the boundary value $W(x_0)$ is always on an exact solution of
 Eq.(2).
 $W(x_0)$ may be also expanded in a power series of $\eps$
\beq
W(x_0)=\sum_{n=0}^{\infty}\eps^n W_n(x_0)
\eeq

If we stop at, say, the $N$-th order, we will have
$\sum_{n=0}^{N}\eps^n z_n(x)\equiv \psi^{(N)}(x; x_0)$
 which is valid only locally at $x\sim x_0$.
However, one may take another point of view as follows:
 Geometrically, we have a family of curves $\{\psi^{(N)}(x; x_0)\}_{x_0}$
 parametrized with $x_0$, and each curve of the family is a good approximation
 around $x=x_0$. Then, if each curve is continued smoothly, the
 resultant curve will valid in a global domain of $x$.
 This is nothing else than to construct the envelope of the family of
 curves.
 More specifically, we only have to determine 
the boundary values $W_n(x_0)$  so that the perturbative solutions 
around $x=x_0$ form an envelope. This is the basic strategy of the RG method
 described geometrically.
Furthermore, to be as accurate as possible, the lowest value 
$W_0(x_0)$ should approximate the exact value $\psi(x_0, x_0)$ as close
 as possible, or $W_{n\geq 1}(x_0)$ should be made as small as possible.

Let us perform the above program.
First, we note that the lowest order solution may be written as
\beq
z_0(x; x_0)=A(x_0)e^{-x^2/4},
\eeq
and $E_0=1/2$; we have made it explicit that 
the amplitude $A(x_0)$ may be dependent on $x_0$.
The choice of the lowest order solution implies that we have also chosen the 
boundary value as
\beq
W_0(x_0)=A(x_0)e^{-x_0^2/4}.
\eeq

The higher order terms with $n\geq 1$ may be  written as
\beq
z_n(x; x_0)=A(x_0)\ee Q_n(x; x_0),
\eeq
where $Q_n(x; x_0)$ is a polynomial of $x$, dependent on $x_0$.
 It is readily shown that the polynomials 
satisfy the same recursion relation Eq.(8) as $P_n(x)$.
However, since we want to make  the boundary value $W_0(x_0)$ as close 
 to the exact one as possible, we  impose the boundary condition as 
\beq
z_{n}(x_0; x_0)=W_n(x_0)=0 \ \ \ {\rm or}\ \ \  
Q_{n}(x_0; x_0)=0,
\eeq
 for $n\geq 1$.

\newcommand{\calP}{Q}
Since $\calP _1(x;x_0)$ satisfies the same equation as $P_1(x)$ does,
one readily obtains 
\beq
\calP _1(x; x_0)= P_1(x) -P_1(x_0),
\eeq
which satisfies the BC Eq.(18).
Notice that a constant is the solution of the homogeneous equation.
The second order equation now reads
\beq
\calP_2''(x;x_0) -x \calP_2'(x;x_0)&=&
 \bigl(\frac{x^4}{4}P_1(x)-\sum_{j=0}^{2}P_{j}(x)E_{n-j}\bigl) \nonumber \\ 
 \ \ & & -P_1(x_0)(\frac{x^4}{4}-E_1). 
\eeq
One can verify that $E_2$ is given Eq.(9).
Since Eq.(20) is linear and the inhomogeneous part
is a linear combination of those for $P_2(x)$ and 
 $P_1(x)$,  $\calP_2(x; x_0)$ is  given by a linear combination of
 $P_2(x)$ and $P_1(x)$;
\beq
\calP _2(x; x_0)= (P_2(x) - P_2(x_0))-P_1(x_0)(P_1(x) -P_1(x_0)),
\eeq
which  satisfies the BC $\calP_2(x_0; x_0)=0$.
One finds that $\calP_3(x;x_0)$ satisfies 
\beq
\calP_3''(x;x_0)&-& x \calP_3'(x;x_0)= 
\{\frac{x^4}{4}P_{2}(x) -\sum_{j=0}^{3}P_{j}(x)E_{n-j}\}\nonumber \\ 
 \ \ \ & & -P_1(x_0)\{\frac{x^4}{4}P_1(x)-\sum_{j=0}^{2}P_{j}(x)E_{n-j}\}
\nonumber \\ 
 \  \ \ & & -(P_2(x_0)-P_1(x_0)^2)(\frac{x^4}{4}-E_1).
\eeq
One can see that the inhomogeneous part is again 
 composed of a linear combination of those for $P_n(x)$ ($n=1, 2, 3$).
Thus the solution satisfying the BC is found to be
\beq
\calP _3(x; x_0)&=& P_3(x) -P_3(x_0) - P_1(x_0)(P_2(x) -P_2(x_0))\nonumber \\
 \ \ \  & & -(P_2(x_0)- P_1(x_0)^2)(P_1(x)- P_1(x_0)),
\eeq
We remark that $E_3$ is the same as what obtained for $P_3(x)$.

Repeating the procedure,
one finds that $Q_n(x; x_0)$ are expressed in terms of 
$P_j(x)$ ($j\leq n$).  For instance,
\beq
\calP_4(x; x_0)&=& P_4(x)-P_4(x_0)-P_1(x_0)(P_3(x)-P_3(x_0))\nonumber \\ 
 \ \ \  & & -(P_2(x_0)-P_1(x_0)^2)(P_2(x)-P_2(x_0))\nonumber \\ 
\ \ \ & & -(P_3(x_0)-2P_1(x_0)P_2(x_0)+P_1(x_0)^3)\nonumber \\  
\ \ \ & &  \cdot (P_1(x)-P_1(x_0)).
\eeq

Thus we obtain the approximate solution valid around $x\sim x_0$
\beq
\psi (x; x_0)\sim A(x_0)\ee \sum _{n=0}^{\infty}\eps^n \calP _n(x; x_0),
\eeq
which satisfies the boundary condition
\beq
\psi(x_0; x_0)=W_0(x_0)=A(x_0)e^{-x_0^2/4}.
\eeq

 One may say that now we have obtained a family of 
 curves $\{\psi(x;x_0)\}_{x_0}$ with $x_0$ parametrizing the curves.
 If $x_0<x<x_0+\Delta x$ with $\Delta x$ being sufficiently small,
 the wave functions $\psi(x; x_0)$ and $\psi(x; x_0+\Delta x)$
 should give the same value at $x$, i.e.,
\beq
\psi(x; x_0)=\psi(x; x_0+\Delta x).
\eeq
 Taking the limit $\Delta x \rightarrow 0$
 this condition is found to yield that
\beq
\frac{d\psi (x; x_0)}{dx_0}\Bigl\vert _{x_0=x}=0.
\eeq
Notice that when $\Delta x \rightarrow 0$, $x \rightarrow x_0$. 
This is the basic equation of our method. This is nothing but the condition
to construct the {\em envelope} of the perturbative wave functions 
valid around $x\sim x_0$. 
It is apparent that the equation has the same form as
 the renormalization group equation, hence the name of the RG method.
The equation gives a condition which $A(x_0)$ must satisfy;
\beq
\frac{dA}{dx}= A(x)\frac{d}{dx_0}\sum _{n=0}^{\infty}
\eps^n(- \calP _n(x; x_0))
\Bigl \vert _{x_0=x}.
\eeq
Defining $f_n(x)$ by 
\beq
-\frac{d}{dx_0}\calP _n(x; x_0)\bigl \vert _{x_0=x}= 
\frac{df_n(x)}{dx},
\eeq 
one obtains 
\beq
A(x)=\bar {A}\cdot {\rm exp}[\sum _{n=0}^{\infty}\eps ^n f_n(x)].
\eeq 
With this solution, the global solution  $\psi _{E}(x)$ is given 
by the boundary value by construction;
\beq
\psi _{E}(x)= W_0(x)&=& A(x)\ee \nonumber \\ 
 \ \ \ &=& 
\bar {A}\cdot {\rm exp}[-\frac{x^2}{4} +\sum _{n=1}^{\infty}\eps ^n f_n(x)] .
\eeq
This is one of the main results of the present paper.

$f_n(x)$'s are easily calculated in terms of $P_n(x)$, and 
 we have
\beq
f_1(x)&=& P_1(x)=-\frac{3}{8}x^2-\frac{1}{16}x^4, \nonumber \\ 
f_2(x)&=& P_2(x)-\frac{1}{2}P_1(x)^2=  \frac{21}{16}x^2+\frac{11}{64}x^4
 +\frac{1}{96}x^6,\nonumber   \\ 
f_3(x)&=&P_3(x)-P_1(x)P_2(x)+\frac{1}{3}P_1(x)^3 , \nonumber \\
 \ \ &=& - \frac{333}{32}x^2-\frac{45}{32}x^4-
  \frac{21}{192}x^6-\frac{1}{256}x^8,\nonumber \\ 
f_4(x)&=&P_4(x)-P_1(x)P_2(x)-\frac{1}{2}P_2(x)^2+P_1(x)^2P_2(x)\nonumber \\ 
 \ \ \ & &  -\frac{1}{4}P_1(x)^4,\nonumber \\ 
\ \ &=&  \frac{30885}{256}x^2 + \frac{8669}{512}x^4 + \frac{1159}{768}x^6  
       + \frac{163}{2048}x^8 + \frac{x^{10}}{512},\nonumber \\ 
f_5(x)&=&P_5(x) -P_1(x) P_4(x) - P_2(x) P_3(x)+P_1(x) P_2(x)^2  \nonumber \\ 
 \ \ \ & &  -P_1(x)^3 P_2(x) + P_1(x)^2 P_3(x) + \frac{1}{5} P_1(x)^5, 
\nonumber \\ 
 \ \ &=& -\frac{916731}{512}x^2 - \frac{33171}{128}x^4 - \frac{6453}{256}x^6 
  \nonumber \\ 
 \ \ \ & & -\frac{823}{512}x^8 - \frac{319}{5120}x^{10} - \frac{7}{6144}x^{12},
 \nonumber \\ 
f_6(x) &=&P_6(x) -P_1(x) P_5(x) - P_2(x) P_4(x)+P_1(x)^2 P_4(x)\nonumber \\  
 \ \ \ & & -\frac{1}{2} P_3(x)^2 + 2 P_1(x) P_2(x) P_3(x) -P_1(x)^3 P_3(x)
  \nonumber \\ 
 \ \ \ & & +\frac{1}{3} P_2(x)^3 - \frac{3}{2} P_1(x)^2 P_2(x)^2 + 
      P_1(x)^4 P_2(x)\nonumber \\ 
 \ \ \ & & -\frac{1}{6} P_1(x)^6,\nonumber \\ 
 \ \ &=& \frac{65518401}{2048}x^2 + \frac{19425763}{4096}x^4 +
  \frac{752825}{1536}x^6 \nonumber \\ 
 \ \ \ & & +\frac{43783}{4096}x^8 + \frac{3481}{2048}x^{10} + 
 \frac{1255}{24576}x^{12} + \frac{3}{4096}x^{14}, 
\eeq
 and so on.
$f_1(x)\sim f_3(x)$ coincide with the results in \cite{bb}, where
explicit expressions of $f_n(x)$ are given  only for $n\leq 3$.
 It is interesting that the polynomials $f_n(x)$ are given in terms of 
 $P_n(x)$  appearing in the naive perturbative expansion in a closed 
form.

 Here learned readers may have suspected that $f_i(x)$'s 
($i=1, 2, 3, ...$) are the {\em cumulant}\cite{cumulant} of the sum 
$\sum_{n=0}^{\infty}\eps^nP_n(x)$ in the sense that 
\beq
\sum_{n=0}^{\infty}\eps^nP_n(x)\sim {\rm exp}
[\sum_{n=0}^{\infty}\eps ^nf_n(x)].
\eeq
In fact, this is the case.
When a function $C(\xi)$ of $\xi$ is given by 
$C(\xi)= \sum _{n=0}^{\infty}\xi ^n/n!\cdot\mu _n$,
the $n$-th cumulant $\lambda_n$ is defined by
$\ln C(\xi) = \sum _{n=0}^{\infty}\xi ^n/n!\cdot\lambda _n$.
Expanding $\ln C(\xi)= \ln (1 + \xi \mu_1 + \xi^2/2\cdot \mu_2 + ...)$,
 one finds that
$\lambda_1=\mu_1,$ \ 
$\lambda_2= \mu_2 -\mu_1^2,$\ 
$\lambda_3 = \mu_3-3\mu_1\mu_2 + 2\mu_1^2,$\  
$\lambda_4= \mu_4-4\mu_1\mu_3 -3\mu_2^2+12\mu_1^2\mu_2 -6\mu_1^4,$\ 
$\lambda_5 =24\mu_1^5 - 60\mu_1^3\mu_2 + 30\mu_1\mu_2^2 + 20\mu_1^2\mu_3$
$ - 10\mu_2\mu_3 -  5\mu_1\mu_4 + \mu_5,$\ 
$\lambda_6=-120\mu_1^6 + 360\mu_1^4\mu_2 - 270\mu_1^2\mu_2^2$
$+ 30\mu_2^3 - 120\mu_1^3\mu_3 + 120\mu_1\mu_2\mu_3$
$- 10\mu_3^2 + 30\mu_1^2\mu_4 - 15\mu_2\mu_4 - 6\mu_1\mu_5 + \mu_6$,
  and so on. Putting $\mu_n=n!P_n$ and $\lambda_n=n!f_n$, one sees that the
 relation Eq. (33) between $P_n$ and $f_n$ is reproduced.
In short, the RG method based on the construction of an envelope
 certainly  resumes the perturbation series of the 
 wave function and the resultant expression are given 
 in terms of  the cumulants of the naive perturbation series. 

Now let us examine  how the
 WKB result Eq.(3) can be constructed from the perturbation series
 obtained above.
 Bender and Bettencourt found that if all terms beyond 
 $1/512\cdot \eps^4 x^{10}$ are neglected,
 the sum of the highest power terms in 
 $f_j(x)$ $(j\leq 4)$ is nicely rewritten as 
\beq
-\frac{x^2}{4}\bigl(1+2\eps x^2+\frac{17}{12}\eps^2x^4+\frac{5}{12}\eps^3 x^6
 +\frac{47}{1152}\eps^4 x^8\bigl) ^{1/8},
\eeq
which behaves for large $x$ as 
\beq
-\sqrt{\eps}\vert x\vert ^3/4(1152/47)^{1/8} \simeq
\sqrt{\eps}\vert x\vert ^3/5.96663
\eeq
 in an excellent agreement with the WKB result.
How about the higher orders.
In the fifth order, the sum of the highest powers may be rewritten 
by neglecting all terms beyond $7\eps^5 x^{10}/1286$ as
\beq 
 \ & & -\frac{x^2}{4}\cdot(1+ \eps x^2/4 -\eps^2 x^4/24 + \eps ^3 x^6/64
-\eps ^4 x^8/128 \nonumber \\ 
 \ & & + 7\eps^5 x^{10}/1286)\nonumber \\ 
 \  &\sim& -\frac{x^2}{4}\bigl(1+\frac{5}{2}\eps x^2 +\frac{115}{48}\eps^2 x^4
 + \frac{35}{32}\eps^3 x^6+ 
\frac{15}{64}\eps^4 x^8\nonumber \\ 
 \ & &  + \frac{4459}{164608}\eps^5 x^{10}\bigl)^{1/10}.
\eeq
For large $x$, the coefficient of $-\sqrt{\eps}\vert x\vert^3$ is
\beq
4(164608/4459)^{1/10}\simeq 5.73827,
\eeq
 which deviates from 6 more 
 than the fourth order result. The sixth order becomes worse:
The sum of the highest powers is rewritten as
\beq
 \ & & -\frac{x^2}{4}\bigl(1+3\eps x^2 +\frac{29}{8}\eps^2 x^4
 + \frac{9}{4}\eps^3 x^6+\frac{577}{768}\eps^4 x^8 \nonumber \\ 
 \ & & +\frac{67621}{493824}\eps^5 x^{10} +\frac{1324349}{35555328}\eps^6x^{12}
\bigl)^{1/12},
\eeq
which makes the coefficient of $-\sqrt{\eps}\vert x\vert^3$ for large $x$ 
\beq
4(35555328/1324349)^{1/12}\simeq 5.26181.
\eeq

In summary, we have successfully  applied the RG method  as 
 formulated in \cite{kuni}
 to Schr\"{o}dinger equation of the quantum anharmonic oscillator (AHO):
 The naive perturbation series of the wave function are 
resummed by the RG equation.
 We have seen that the resummation is performed in 
 the RG method more
 mechanically and explicitly than in MSPT.
 We have shown that the resummed series $\sum_{n=0}^{\infty}\eps^nf_n(x)$ is
 the {\em cumulant} 
 of the naive perturbation series.
We have worked out up to the sixth order in the perturbative
 expansion 
 and found the following:
Although the sum of the highest power in $f_n(x)$
 can be organized so that it becomes asymptotically  proportional
 to $\sqrt{\eps}\vert x\vert^3$ as was done  in \cite{bb,gm}, 
the coefficient of it  
 reaches the  closest value  to 6, the WKB result, in the fourth order,
 then  goes away monotonously from the closest value in the higher
 orders. This is plausible because the convergence radius of the
 perturbation series is zero; the cumulant series should be  at best  
 an asymptotic series. 

We remark that the RG method as developed here can be also applied to
 the first excited state; it is, however, unlikely that the method can be 
 used to the higher excited states beyond the first excited state.

Finally, we mention that a variational perturbation 
 method called the delta-expansion method\cite{delta} has been 
 extended for obtaining wave functions\cite{prl}. 
 The key ingredient of the extension is to
construct an envelope of a set of 
 perturbative wave functions as in the RG method but 
{\em with a variational parameter}.
In this method, although 
the basic equation  can not be solved analytically but
 only numerically, 
  uniformly valid wave functions  with correct asymptotic behavior
 are obtained in the first-order 
 perturbation even for strong couplings and for excited states.
In the present method, the basic equations are solved analytically,
 and the asymptotic form of the  wave function is constructed explicitly,
 although a further resummation devised  in \cite{bb,gm} 
is needed for obtaining the asymptotic
 form. In this sense, the two methods are complementary.

This work was partially supported by the Grants-in-Aid of the Japanese Ministry
 of Education, Science and Culture (No. 09640377).

\end{document}